\documentclass[a4paper,twoside,reqno]{bjp}
\usepackage{graphicx}
\usepackage{cite}
\usepackage{amssymb,amsmath,amscd,amsthm}
\usepackage{times}

\usepackage[bookmarks=false]{hyperref}
\hypersetup{%
    colorlinks=true,        % false: boxed links; true: colored links
    linkcolor=blue,          % color of internal links (change box color with linkbordercolor)
    citecolor=blue,         % color of links to bibliography
    urlcolor=blue           % color of external links
    }

\usepackage{geometry}
\allowdisplaybreaks

\begin{document}

\title{Clifford Solver for the Tetrahedron Equation and its Variants}

\runningheads{P.Padmanabhan {\it et al.}}{Clifford Algebras and Tetrahedron Equations}

\begin{start}{
\author{P.Padmanabhan}{1},
\author{V.K.Singh}{2},
\author{V.E.Korepin}{3}

\address{Department of Physics, School of Basic Sciences,\\ Indian Institute of Technology, Bhubaneswar, 752050, India}{1}
\address{Center for Quantum and Topological Systems (CQTS), \\ NYUAD Research Institute, New York University Abu Dhabi, 129188, Abu Dhabi, UAE}{2}
\address{C. N. Yang Institute for Theoretical Physics, \\ Stony Brook University, New York 11794, USA}{3}

\received{1 November 2025}
}

\begin{Abstract}
The different forms of the tetrahedron equation appear when all possible ways to label the scattering process of infinitely long straight lines are considered in three dimensional spacetime. This is expected to lead to three dimensional integrability, analogous to the Yang-Baxter equation. Among the three possibilities, we consider two of them and their variants. We show that Clifford algebras solve both the constant and the spectral parameter dependent versions of all of them. We also present a scheme for canonically solving higher simplex equations using tetrahedron solutions. 
\end{Abstract}

\begin{KEY}
Clifford algebras, Higher dimensional integrability, Tetrahedron equations, Simplex equations
\end{KEY}
\end{start}
%%%%%%%%%%%%%%%%%%%%%%%%%%%

%%%%%%%%%%%%%%%%%%%%%%%%%%%%%%
\section{Introduction}
\label{sec:introduction}
%%%%%%%%%%%%%%%%%%%%%%%%%%%%%%%
The Yang-Baxter equation \cite{cny,bax2} is pivotal for its role in the construction and analysis of integrable models in (1+1) dimensions using the algebraic Bethe ansatz \cite{QISM-book}. Its three  dimensional generalization, known as the tetrahedron equation, was formulated in the early 80's by Zamolodchikov \cite{z2}. Early solutions were studied by Baxter \cite{bax1} and Bazhanov \cite{baz1} but were found insufficient for physical applications in spite of their ingenuity. Finding more and interesting solutions to this equation would lead to new developments in exactly solvable models and in higher dimensional algebras. Thus it is important to find universal methods that lead to solutions of the tetrahedron and the higher dimensional counterparts known as higher simplex equations.

In this paper we write down one such universal method using Clifford algebras. Ans\"{a}tze constructed using such algebras help solve the different forms of the tetrahedron equation and its variants. These different forms are motivated through the labeling of scattering process in three dimensional spacetime in Sec. \ref{sec:formulation}. The Clifford solutions make up Sec. \ref{sec:solutions}. Here we also show a canonical approach to construct the solutions of higher simplex equations using a tetrahedron solution. A summary of the findings and a brief outlook is provided in Sec. \ref{sec:conclusions}.

%%%%%%%%%%%%%%%%%%%%%%%%%%%%%%
\section{Formulation}
\label{sec:formulation}
%%%%%%%%%%%%%%%%%%%%%%%%%%%%%%%
The tetrahedron equation can be written down in different forms.
Here we will only show the types that we will solve using algebraic methods. The main source for the several forms is the scattering of three straight lines in two dimensional space. From a three dimensional spacetime perspective this process forms a tetrahedron or a 3-simplex. The different forms correspond to the different labeling describing this scattering process \cite{hie1}. These include:
\begin{enumerate}
    \item {\it vacua - cell or volume labeling} 
    \item {\it string - face labeling}
    \item {\it particle - edge labeling}
\end{enumerate}
In a given time slice, the vacua, string and particle correspond to the face, edge and vertex of the intersecting strings in two dimensional space. Whereas, in the three dimensional spacetime picture they correspond to the volume, face and edge respectively. The number of labels for the scattering matrix elements differ according to the chosen scheme. The different forms are related to each other {\it via} Wu-Kadanoff type of dualities \cite{wu,kad}.

The constant tetrahedron equation corresponding to the particle or vertex form is given by 
\begin{eqnarray}\label{eq:TE-vertex-form-constant}
    R_{ijk}R_{ilm}R_{jlp}R_{kmp} = R_{kmp}R_{jlp}R_{ilm}R_{ijk}.
\end{eqnarray}
Here $R$ denotes the {\it tetrahedron operator}. It acts on $V\otimes V\otimes V$, with $V$ being the local Hilbert space. For finite dimensional matrix solutions we take $V\simeq\mathbb{C}^d$. The spectral parameter dependent version of Eq. (\ref{eq:TE-vertex-form-constant}) is given by 
\begin{eqnarray}\label{eq:TE-vertex-form-spectral}
    & & R_{ijk}(z_{ijk})R_{ilm}(z_{ilm})R_{jlp}(z_{jlp})R_{kmp}(z_{kmp}) \nonumber \\ &  =& R_{kmp}(z_{kmp})R_{jlp}(z_{jlp})R_{ilm}(z_{ilm})R_{ijk}(z_{ijk}),
\end{eqnarray}
with $z_{ijk}$ representing a tuple of complex spectral parameters. Physically we can view $R$ as providing the scattering amplitude between infinitely long straight strings scattering in two dimensional space. The spacetime picture corresponds to world sheets of the scattering strings with the tetrahedron equation implying a factorization condition of the scattering matrix into a product of three world sheets at a time. The latter should be read with the analogous situation for the Yang-Baxter equation in mind. 

We will also solve the string or face labeled tetrahedron equation. We will write down a related form known as the {\it Frenkel-Moore equation} \cite{FM},
\begin{eqnarray}\label{eq:TE-edge-form-constant}
    R_{ijk}R_{ijl}R_{ikl}R_{jkl} = R_{jkl}R_{ikl}R_{ijl}R_{ijk},
\end{eqnarray}
and an equation analogous to Eq. (\ref{eq:TE-vertex-form-spectral}) for the spectral parameter dependent version. This form is related to the original form of the tetrahedron equation \cite{z2}.

Our methods will also solve two other variants of the tetrahedron equation that originate from three dimensional integrability. The first is known as the {\it quantized Yang-Baxter equation} \cite{kunbook}. The constant form of this equation, also referred to as the $RLLL$ form, reads 
\begin{eqnarray}\label{eq:RLLL}
    L_{12a}L_{13b}L_{23c}R_{abc} = R_{abc}L_{23c}L_{13b}L_{12a}.
\end{eqnarray}
Here $a$, $b$ and $c$ index auxiliary spaces. The similarity to the $RLL$ form of the Yang-Baxter equation should be noted. The associativity condition for the algebra generated by the $L$ operators then leads to the vertex form of the tetrahedron equation for $R$ in Eq. (\ref{eq:TE-vertex-form-constant}). The spectral parameter dependent form can be obtained by including the complex argument $z_{ijk}$ to each $L_{ijk}$ and $R_{ijk}$.

Using the notation of \cite{kunbook}, the second and final form of the tetrahedron equation is of the $MMLL$ type. Its constant form is given by 
\begin{eqnarray}\label{eq:MMLL}
    M_{126}M_{346}L_{135}L_{245} = L_{245}L_{135}M_{346}M_{126}.
\end{eqnarray}
The spectral parameter dependent form is generated as mentioned earlier.
Note the difference in the index structures of all the three forms in Eqs. (\ref{eq:TE-vertex-form-constant}, \ref{eq:TE-edge-form-constant}, \ref{eq:MMLL}). Regardless, all of them can be solved using Clifford algebras as we shall now see.

%%%%%%%%%%%%%%%%%%%%%%%%%%%%%%%%%%%
\section{Solutions}
\label{sec:solutions}
%%%%%%%%%%%%%%%%%%%%%%%%%%%%%%%%%%%
Consider a pair of operators $A$ and $B$ that follow the relations
\begin{eqnarray}\label{eq:AB-algebra}
   \{A_i, B_i\} =0~~;~~\left[A_i, B_j\right]=0~\textrm{for}~i\neq j.
\end{eqnarray}
There are no additional constraints on $A$ and $B$. Such operators can be realized using Clifford algebras. 

Using this algebra we find that the following ans\"{a}tze solve the vertex form of the tetrahedron equation, Eq. (\ref{eq:TE-vertex-form-spectral}) and the quantized Yang-Baxter equation, Eq. (\ref{eq:RLLL}):
\begin{eqnarray}\label{eq:20-30-RRRR}
    & R_{ijk}(z_{ijk}) = \alpha_{ijk}~A_iA_jB_k + \beta_{ijk}~A_iB_jA_k + \gamma_{ijk}~B_iA_jA_k + B_iB_jB_k,  & \nonumber \\
    & L_{ijk}(z'_{ijk}) = \alpha'_{ijk}~A_iA_jB_k + \beta'_{ijk}~A_iB_jA_k + \gamma'_{ijk}~B_iA_jA_k + B_iB_jB_k.  &
\end{eqnarray}
The complex parameters $\alpha$, $\beta$ and $\gamma$ are arbitrary implying that the above ansatz provides an entire family of tetrahedron solutions.
Another class of solutions is obtained by interchanging $A$ and $B$ in the above expression. The analytic proof that this indeed solves the vertex form of the spectral parameter dependent tetrahedron equation, Eq. (\ref{eq:TE-vertex-form-spectral}), is shown in \cite{pp1}. The solution to the constant form, Eq. (\ref{eq:TE-vertex-form-constant}), is obtained when we suppress the $ijk$ indices on the spectral parameters. The family of solutions in Eq. (\ref{eq:20-30-RRRR}) also solve the constant and spectral dependent forms of the Frenkel-Moore equation, Eq. (\ref{eq:TE-edge-form-constant}).

To solve the $MMLL$ type of the tetrahedron equation, Eq. (\ref{eq:MMLL}), we use the ans\"{a}tze:
\begin{eqnarray}
    & M_{ijk}(z_{ijk}) = \alpha_{ijk}~A_iA_jB_k + \beta_{ijk}~A_iB_jA_k + \gamma_{ijk}~B_iA_jA_k + B_iB_jB_k, & \nonumber \\
     & L_{ijk}(z_{ijk}) = \alpha_{ijk}~B_iB_jA_k + \beta_{ijk}~B_iA_jB_k + \gamma_{ijk}~A_iB_jB_k + A_iA_jA_k.  &
\end{eqnarray}
Interchanging the ans\"{a}tze for the $M$ and $L$ operators provide another set of solutions to Eq. (\ref{eq:MMLL}).

We will now present a canonical way to solve $n$-simplex equations, first formulated in \cite{baz1,ca-sa,Mai1}, using tetrahedron solutions. This is demonstrated for the 4-simplex equation \cite{baz1}, which then points to the generalization to higher $n$. We find that the 4-simplex operator decomposed into a product of tetrahedron operators with spectral parameters in the following way,
\begin{eqnarray}\label{eq:4-simplex-T}
   & & R_{\{1234\}}(u,v,w,t)  \nonumber \\  & = & T_{123}(v,w,t)T_{1'2'4}(u,w,t)T_{1''3'4'}(u,v,t)T_{2''3''4''}(u,v,w),
\end{eqnarray}
satisfies the 4-simplex equation
\begin{eqnarray}
    & & R_{\{1234\}}(u,v,w,t)R_{\{1567\}}(u,v,w,s)R_{\{2589\}}(u,v,t,s)  \nonumber \\ &\times & R_{\{368,10\}}(u,w,t,s)R_{\{479,10\}}(v,w,t,s) \nonumber \\
    & = & \left[\leftrightarrow \textrm{reverse} \right],
\end{eqnarray}
when the tetrahedron operators $T$ satisfy
\begin{eqnarray}
    & & T_{ijk}(u,v,w)T_{ilm}(u,v,t)T_{jlp}(u,w,t)T_{kmp}(v,w,t) \nonumber \\
    & = & \left[\leftrightarrow \textrm{reverse} \right].
\end{eqnarray}
We have used a shorthand notation for the indices in the braces $\{ \cdot\}$,
\begin{eqnarray}
    \{ijkl\} = ii'i'', jj'j'', kk'k'', ll'l''.
\end{eqnarray}
Note that this holds for any valid tetrahedron solution requiring no further constraints on $T$. This shows that the decomposition in Eq. (\ref{eq:4-simplex-T}) naturally solves the 4-simplex equation in an appropriately `enlarged' Hilbert space. A similar method to obtain tetrahedron solutions from 2-simplex operators is shown in \cite{Mai2}.

This procedure can be easily generalized to higher simplex equations. To do this we only need to compute the enlarged dimension of each index of the $n$-simplex equation, such that its index structure can accommodate multiple tetrahedron equations. For instance when $n=5$, each index of the 5-simplex operator will need to split into a minimum of 12 different indices giving us a total of 60 indices in each 5-simplex operator. This then implies that each 5-simplex operator decomposes into 20 tetrahedron operators. Thus finding the minimum number of indices into which a given index of the $n$-simplex operator splits, suggests the number of the tetrahedron operators the $n$-simplex operator decomposes into. This is given by the formula
\begin{eqnarray}
    n\times p = 3\times 4m~;~m,p\in\mathbb{Z}^+.
\end{eqnarray}
The minimum $m$ gives the minimum value of the integer $p$ when this equation is satisfied.

%%%%%%%%%%%%%%%%%%%%%%%%%%%%%%%%%%%
\section{Conclusions} 
\label{sec:conclusions}
%%%%%%%%%%%%%%%%%%%%%%%%%%%%%%%%%%%
In this short note we have summarized two types of the tetrahedron equation derived from constraints of the scattering processes of straight lines in three dimensional spacetime. These are the vertex form, Eq. (\ref{eq:TE-vertex-form-spectral}) and the edge form, Eq. (\ref{eq:TE-edge-form-constant}). Their variants, Eqs. (\ref{eq:RLLL}, \ref{eq:MMLL}), motivated from three dimensional integrability, were also considered. An entire family of solutions to all these different equations were constructed using Clifford algebras. For more details, including analytical proofs and explicit matrix representations, we refer the reader to \cite{pp1}. The techniques outlined in this paper also extend to higher simplex equations and to the higher dimensional reflection equations \cite{kunbook}. It is important to note that several of the solutions shown in this work have positive Boltzmann weights. Thus they can find potential application in the  construction of three dimensional statistical physics models. More algebraic techniques, such as Majorana fermions, partition algebras among others, for solving the Yang-Baxter, tetrahedron and other higher simplex equations can be found in \cite{pp2,pp3,pp4,pp5}.

%\section*{Appendix} An appendix, if needed, appears before the acknowledgements.

\section*{Acknowledgements} 
VEK is funded by the U.S. Department of Energy, Office of Science, National Quantum Information Science Research Centers, Co-Design Center for Quantum Advantage ($C^2QA$) under Contract No. DE-SC0012704.  The work of VKS is supported by ``Tamkeen under the NYU Abu Dhabi Research Institute grant CG008''.

\end{document}